\documentclass[aps,twocolumn,amsmath,amssymb,showpacs]{revtex4}  
\usepackage{graphicx}  
\usepackage{dcolumn}   
\usepackage{bm}        
\usepackage{amssymb}   

\begin{document}

\title{Alternative description of particle shower longitudinal profile}

\author{Samvel Ter-Antonyan}
\altaffiliation{}\email{samvel@icecube.wisc.edu}\affiliation{Department of Physics, Southern University, USA}
\date{\today}

\begin{abstract}
An alternative parametrization of particle shower longitudinal profile is presented. 
The accuracy of obtained shower profile description is about $2-3\%$  for the  $0-1500$ g/cm$^2$ 
atmosphere slant depths and primary H, He,$\dots$Fe nuclei in $1$ PeV-$10$ EeV energy range. 
It is shown that the shape of shower profile depends only on the nucleon energy, whereas
the maximum shower size also depends on the energy of parental  nucleus.
Results are based on the {\footnotesize{CORSIKA}} simulated shower profiles and are presented in comparison
with Gaisser-Hillas parametrization.
\end{abstract}
\pacs{96.50.S-, 96.50.sd, 96.50.sb}
\maketitle
\section{INTRODUCTION} 
The parametrization of longitudinal profiles for particle showers produced by the primary
nuclei in atmosphere, is an essential tool for the primary nuclei identification 
and the evaluation of primary energy. Experiments that sample the shower
longitudinal development using Cherenkov light images \cite{HiRes} or air fluorescence
\cite{Auger,TArray} from different traversed atmosphere depths, extract the position of shower maximum,
which is sensitive to the incident primary nucleus.   
The integral of shower profile strongly correlates with primary energy \cite{Matthews}.

The shower longitudinal profile is a dependence  of the shower particle number 
($N$) on a given traversed atmospheric depth, $T$. The parametrization of shower profile commonly
used in cosmic-ray experiments is Gaisser-Hillas formula \cite{GH}:  
\begin{equation}\label{one}
N(X)=N_{max}\left(\frac{X}{X_{max}}\right)^{X_{max}}\exp{(X_{max}-X)}\;,
\end{equation}
where $X=(T-X_0)/\lambda$ and $X_{max}=(T_{max}-X_0)/\lambda$.

The maximum number of shower particles $N_{max}$ at the traversed atmosphere depth $T_{max}$ along with $X_0$ and $\lambda$ 
in expression (1) are free parameters that depend on the primary nucleus and energy.

The standard primary nuclei composition consists of
the first 28 nuclei of the periodic table with mass (nucleon) numbers $A=1,\dots56$
usually divided into four-six groups (species) H, He, CNO-like, Si-like, Fe-like. 
The large number of nuclei species (more than four) increases the uncertainties of the
inverse problem ($E$ and $A$ reconstruction) falsely improving the agreement of experiment
with the theory \cite{samo}.   

The primary energy region responsible for particle shower 
detection at the observation level begins at about $E>1$ PeV and ends at GZK cutoff
energies \cite{Auger}. 

The efficiency of four-parametric parametrization (1) is in its 
applicability to a wide range of energies and primary nuclei. 
However, the observed correlations between parameters 
result in a loss of the physical meaning of $X_0$ and $\lambda$ \cite{Mont} and reduce the range 
of effective atmosphere depths for Eq.~(1).
\section{Parametrization}  
Here, an alternative parametrization $N(T,E,\boldsymbol{\varepsilon})$ for particle shower longitudinal profile is proposed
using three non-correlating parameters that depend on the primary particle energy and
nucleon energy, $\boldsymbol{\varepsilon}$:  
\begin{equation}\label{two}
N(x)=N_{\max}\exp{(-\frac{1}{2}\left(\frac{\ln{x}}{\delta(x)}\right)^2)}\;,
\end{equation}
where
\begin{equation}\label{three}
\delta(x)=\alpha-\beta(\tanh{x})^{\frac{1}{4}}
\end{equation}
is the profile shape function of variable 
\[
 x=\frac{T}{T_{max}}\;.
\]

The shower maximum position, $T_{max}(\boldsymbol{\varepsilon})$, and 
shape function, $\delta(x,\boldsymbol{\varepsilon})$, turned out to be 
 dependent on the primary particle energy per nucleon,
 \[
{\boldsymbol{\varepsilon}}=\frac{E}{A}\;,\;\;(\text{PeV/n}).
\]
The maximum number of shower particles, $N_{max}(E,\boldsymbol{\varepsilon})$
is factored into the primary energy and a function of nucleon energy only. The corresponding approximations 
for the parameters of shower longitudinal profile (2,3) are: 
\begin{subequations}
\begin{align}
&\alpha =0.707+0.209{\boldsymbol{\varepsilon}}^{-0.084}\;,\;\;\;\;\; \beta =\sqrt{\alpha/2.59}\;,\\
&T_{max} =433.5+38.9(\ln({\boldsymbol{\varepsilon}}A_{Fe}))^{0.857},\;\text{(g/cm$^2$)},\\
&N_{max} =0.653(E/1\text{GeV})(1-e^{-2.5{\boldsymbol{\varepsilon}}^{0.12}})\;,
\end{align}
\end{subequations}
 where $A_{Fe}=56$ and $\boldsymbol{\varepsilon}$ is in the units of PeV/n. The goodness-of-fit tests for (4a-c) were $\chi^2<1$ at
 negligible correlations between the $\alpha$, $T_{max}$ and $N_{max}$ parameters.
\section{Shower Profiles}  
The values of 
free parameters in expressions (1) and (2) were obtained from simulated shower profiles
(training sample) using {\footnotesize{CORSIKA}} \cite{CORSIKA} ({\footnotesize{SIBYLL}} \cite{SIBYLL}) code for four primary nuclei $A\equiv1, 4, 16, 56$
at six energies $E\equiv1, 10, 100, 500, 2500, 10^4$ PeV. Shower profiles were studied for
10 atmosphere depths $T\equiv 100, 200,\dots1000$ g/cm$^2$ at two zenith angles, $\cos\theta=0.7$ and $1$.
The shower particle energy threshold was $E_e>1$ MeV. Simulation statistics were provided for less than 2-3\%
statistical errors in the whole measurement range.  

\begin{figure}
\includegraphics[scale=0.35]{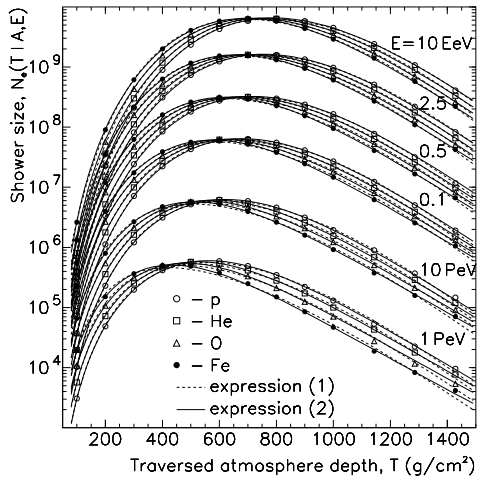}
\caption{\label{curve1} Average longitudinal shower profiles for 13 traversed atmosphere depths  
produced by H, He, O and Fe primary nuclei with 6 energies (from 1 PeV to 10 EeV).
The symbols are {\footnotesize{CORSIKA}} shower simulated data (training sample).
The dashed lines are results from the 4-parametric approximation (1). The solid lines  
are parametrizations (2-4) computed for corresponding primary nuclei and energies.}
\end{figure}

The averaged shower profiles were approximated by expressions (1) and (2) using 13 reference depths.
The results are presented in Fig.~\ref{curve1}. It is seen that the parametrization (1) (dashed lines) underestimates the shower sizes
at large  atmosphere depths.

\begin{figure}
\includegraphics[scale=0.35]{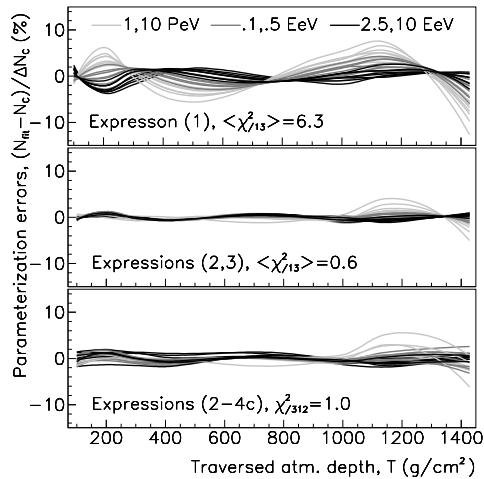}
\caption{\label{err}  Parameterization errors $(N_{fit}-N_{corsika})/\Delta N_{corsika}$ for the 4-parametric expressions (1) and (2,3) 
are shown in the upper
and middle panels respectively. The lower panel corresponds to the errors of  the shower profiles $N(T,A,E)$ from  
expressions (2-4) for different primary energies and nuclei.    
}
\end{figure}

The parametrization errors of expressions (1) and (2) and corresponding $\chi^2_{/\text{d.o.f.}}$ are presented in Fig.~\ref{err}
for different primary energies and nuclei.
The upper and middle panels show the errors of the four-parametric approximations of {\footnotesize{CORSIKA}}
simulated shower profiles using the $N_{max}, T_{max}, X_0$ and $\lambda$ parameters  of expression (1)
and the $N_{max}, T_{max}, \alpha$ and $\beta$ of expressions (2,3). The lower panel of Fig.~\ref{err} shows
the errors of shower profiles $N(T,E,A)$ from expressions (2-4).

\begin{figure}
\includegraphics[scale=0.35]{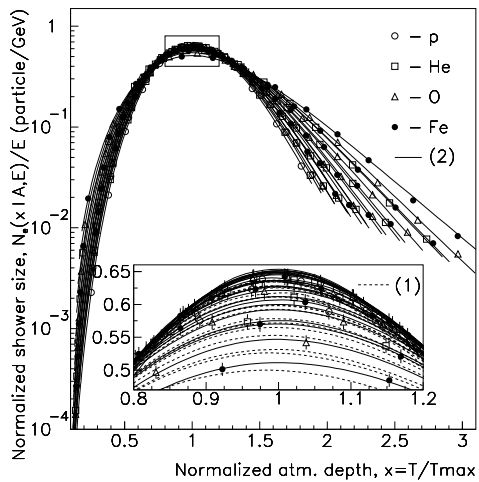}
\caption{\label{curve2}Normalized particle shower profiles. The symbols are {\footnotesize{CORSIKA}} 
simulated data (training sample). The lines are the results of parametrization (2). Inset figure
is a zoom of selected rectangular region for $0.8<x<1.2$.  The dashed lines correspond to 
the parametrization (1).  
}
\end{figure}

The normalized simulated (symbols) and parametrized (lines) shower profiles 
are presented in Fig.~\ref{curve2}. It is seen that the parametrization (2)
effectively describe the shower profiles in the regions of both the maximum ($x\simeq1$, inset figure, solid line) 
and asymptotic depths ($x\simeq3$). Equation (1) is systematically biased about $-2\%$ 
(inset figure, dashed line) at $x\simeq1$.
     
\section{Parameters}   
\begin{figure}
\includegraphics[scale=0.4]{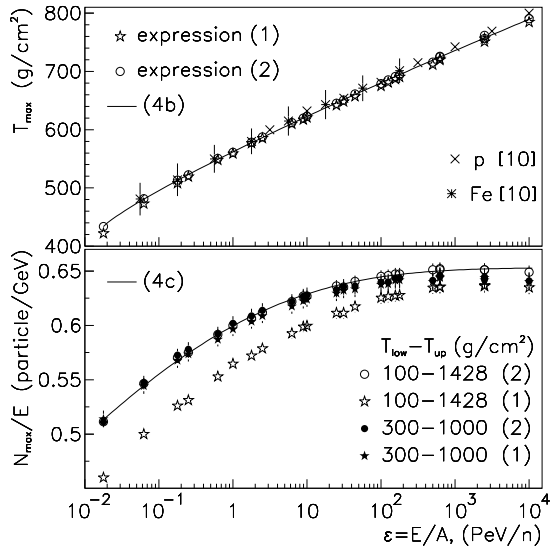}
\caption{\label{Nmax}Parameters $T_{max}$ (upper panel) and $N_{max}/E$ in units of particle/GeV (lower panel)
derived from expressions (1) and (2) for the different boundaries of traversed depths. 
The lines correspond to  the expressions  (4b) and (4c) for $T_{max}$ and $N_{max}$ respectively. 
}
\end{figure}

The study of $T_{max}(\boldsymbol{\varepsilon})$ and $N_{max}(\boldsymbol{\varepsilon})$ dependence 
on nucleon energy ($\boldsymbol{\varepsilon}$) are presented in the upper and lower panel of Fig.~\ref{Nmax}  respectively.
The approximations of shower profiles using
parametrizations (1) and (2) were trailed for different lower ($T_{low}$) and upper ($T_{up}$) limits of traversed
atmosphere depth.

 The estimated values for $T_{max}$ (Fig.~\ref{Nmax}, upper panel) were unbiased for all trails.
 The line in Fig.~\ref{Nmax} corresponds to the expression (4b). The asterisk and cross symbols in Fig.~\ref{Nmax}
 are correspondingly renormalized {\footnotesize{CORSIKA}} simulated data from  \cite{Swordy}.
 
 Estimations of $N_{max}$ (Fig.~\ref{Nmax}, lower panel) using expression (1) for approximations of
 shower profile turned out to be dependent on boundary conditions for atmosphere depth (hollow and bold star symbols),
 whereas the expression (2) remained practically unbiased (hollow and bold circle symbols) for different boundaries. 
 
\begin{figure}
\includegraphics[scale=0.4]{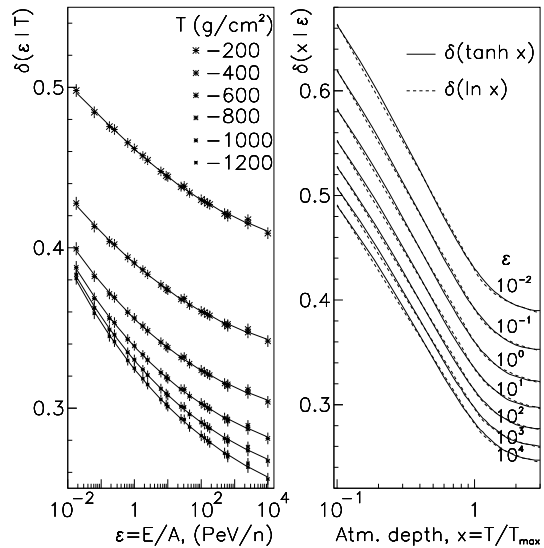}
\caption{\label{delta}Shower profile shape function (expression (3)) depending on the nucleon energy (left panel)
and normalized atmosphere depth (right panel). The solid lines correspond to the expressions (3,4a). The dashed lines in the right panel
are the logarithmic simplifications of the shape function according to (5,6).
}
\end{figure}

The shower profile shape functions $\delta(\boldsymbol{\varepsilon}|T)$ and $\delta(x|\boldsymbol{\varepsilon})$  are presented in Fig.~\ref{delta},
where the symbols (left panel) are the data extracted from {\footnotesize{CORSIKA}} simulated training sample. The solid lines in both panels correspond to the expressions (3,4a).
The dashed lines in the right panel of Fig.~\ref{delta} are the $0.5\%$ accuracy logarithmic simplifications of shape function (3),
\begin{equation}
\delta(x)\simeq
\begin{cases}
a-b\ln{x}, &\text{if $0.07\lesssim x\leqq1$;}\\
a-b\ln{x}/(1+1.59\ln{x}), &\text{if $x\geqq1$,}
\end{cases}
\end{equation}
where 
\begin{equation} 
\begin{split}
a& =0.215+0.145{\boldsymbol{\varepsilon}}^{-0.084}\;,\\
b& =0.086+0.011{\boldsymbol{\varepsilon}}^{-0.084}\;,
\end{split} 
\end{equation}
at $\chi^2/450=0.7$.
The approximation (5) provides the analytic solution of the inverse function $N_0^{-1}(x)$ for $x<1$ and $N_1^{-1}(x)$ for $x>1$. 
\section{Verification} 
\begin{figure}
\includegraphics[scale=0.4]{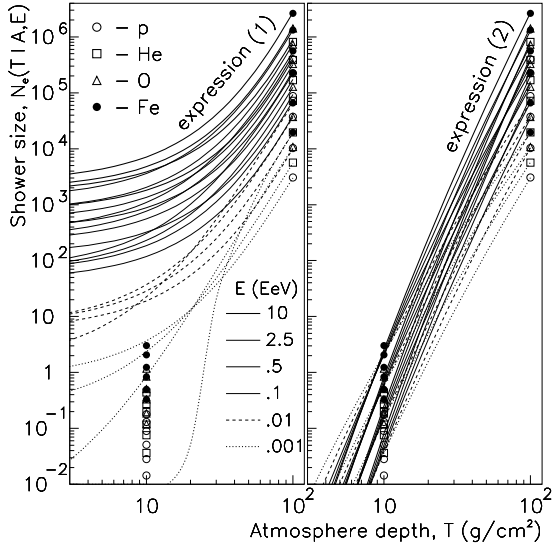}
\caption{\label{zero} Extrapolations of the parametrization (1) (left panel) and (2) (right panel) to the earliest 
stage of shower development for different primary nuclei and energies (lines). The symbols are
the {\footnotesize{CORSIKA}} simulated data.
}
\end{figure}

The verification of the universality of approximations (1) and (2) was performed by extrapolating the shower
profiles from $100-1428$ g/cm$^2$ interval to the $T=10$ g/cm$^2$ observation level,
corresponding to the earlier stage of shower development. The results are presented in Fig.~\ref{zero}. 
The symbols at $T=10$ g/cm$^2$ in Fig.~\ref{zero} are the corresponding data from {\footnotesize{CORSIKA}} 
simulated control sample, whereas  the symbols at $T=100$ g/cm$^2$ are the representatives of the training sample (Section III).

It is seen that parametrization (1) being trained in the $100-1428$ g/cm$^2$ depth interval can not be extrapolated
to the region less than about 50 g/cm$^2$ (lines, left panel), whereas the parametrization (2) works correctly up to the beginning of 
the atmosphere (lines, right panel). 

\begin{figure}
\includegraphics[scale=0.4]{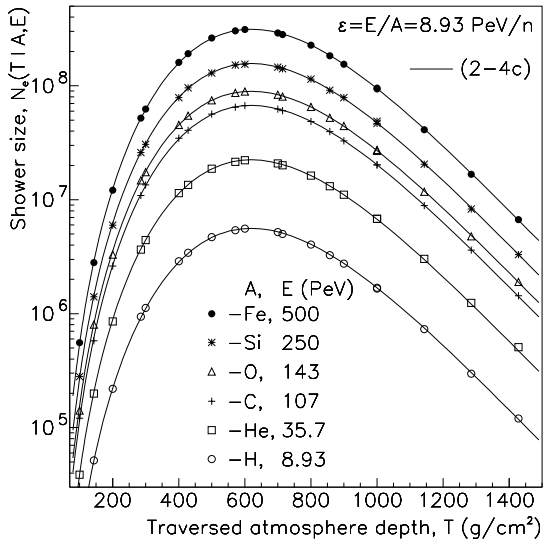}
\caption{\label{verify}Control samples of shower profiles  (symbols) produced by the   
different primary nuclei with the same nucleon energy $\boldsymbol{\varepsilon}\simeq8.93$ PeV/n. 
The lines are the predictions from (2-4).  
}
\end{figure}

The verification of the shower profiles (2-4) by the control samples of different nuclei and energies 
are shown in Fig.~\ref{verify}.
The shower profile for primary Fe nucleus with energy $E=500$ PeV and corresponding $\boldsymbol{\varepsilon}\simeq8.93$ PeV/n
from training sample (Section III) are compared to the control sample of shower profiles produced by primary H, He, C, O and Si nuclei
with the same energy per nucleon (symbols). The lines in Fig.~\ref{verify} are the corresponding congruent predictions 
from the parametrizations (2-4).

The results in Fig.~\ref{verify} confirm the $\boldsymbol{\varepsilon}$-dependence of the shower longitudinal profile shape
(expressions 4a,b). The shower profile amplitude ($N_{max}$) also depends linearly on the primary energy, $E$ (expression 4c).  

The good agreement in Fig.~\ref{verify} between predictions (lines) and simulated data
indicates the correctness of expressions (2-4) for shower profile description at least
with the accuracies of about $2-3\%$ in the whole measurement range.
\section{\label{sec2} INTEGRAL} 
The right hand side of parametrization (2) at the corresponding normalization can be considered
 as a probability density function and be used for primary energy evaluation \cite{Auger,Matthews}.
Unfortunately this function was missed by mathematicians and
by using numerical technique the required normalization
\begin{equation}\label{int}
\int_0^\infty{f(x,\boldsymbol{\varepsilon})\mathrm{d}x}\simeq1\pm10^{-4}
\end{equation}
was provided for probability density function
\begin{equation}
f(x)=\frac{1}{\sqrt{2\pi}\delta_0}\exp{(-\frac{1}{2}\left(\frac{\ln{x}}{\delta(x)}\right)^2)}
\end{equation}
with additional parameter
\[
\delta_0=0.226+0.148{\boldsymbol{\varepsilon}}^{-0.092}.
\]

The goodness-of-fit test  for  $\delta_0(\boldsymbol{\varepsilon})$ was $\chi^2=0.01$  in the $10^{-2}\leq\boldsymbol{\varepsilon}\leq10^4$ (PeV/nucleon) interval
and the upper limit of integral (7), $x_{max}=3$.  

It is interesting to note the relation between parameters $\delta_0$ and
shape function $\delta(x)$ from expression (3):
\begin{equation}
\delta_0(\boldsymbol{\varepsilon})\simeq\frac{1}{x_{max}}\int_0^{x_{max}}\delta(x)\mathrm{d}x\; \pm1\%\;.
\end{equation} 
The statistical parameters, the average ($\bar{x}$) and  standard deviation ($\sigma_x$) of distribution (8), are well approximated 
($0.1\%$ errors) by the following expressions that depend on the nucleon energy: 
\[
\bar{x}=1.036+0.094{\boldsymbol{\varepsilon}}^{-0.12}
\]
at  $\chi^2=0.1$, and 
\[
\sigma_x=0.226+0.176{\boldsymbol{\varepsilon}}^{-0.092}
\]
at $\chi^2=1.1$.
\section{Fluctuations}
The main source of shower profile fluctuations is the depth of the first interaction of primary
particles in the atmosphere \cite{Stanev}. The exponentially distributed uncertainty of the
first interaction point results in the corresponding fluctuations of the shower profile (2) depending on the rate
of change ($\mathrm{d} N/\mathrm{d} x$) of the profile with respect to the depth, $x$. Thus, the fluctuations
should be maximal at the beginning of shower development ($x\simeq0$, Fig.~\ref{curve2}), 
and minimal in the region of shower maximum, $x=1$. The dependence of the interaction length, $\lambda(A,E)$,
on the primary particle  also results in the mass ($A$) and energy ($E$) dependences of shower profile fluctuations.

\begin{figure}
\includegraphics[scale=0.4]{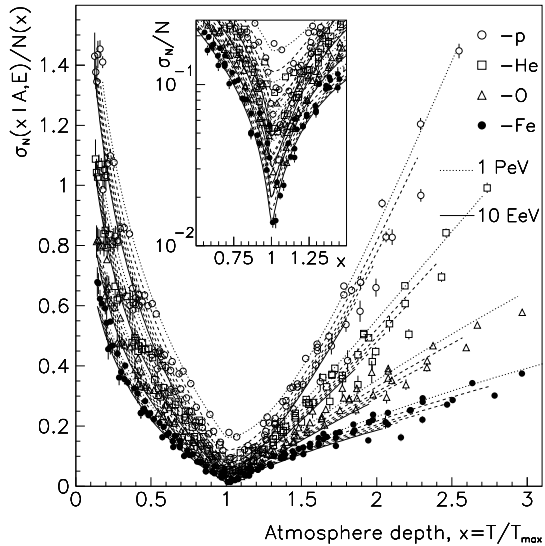}
\caption{\label{sigma} Normalized standard deviations ($\sigma_N/N$) of shower particles for different primary nuclei and primary energies (symbols).  
The lines are represent the parametrization (10) for energies of 1 PeV (dotted lines) and 10 EeV (solid lines). The dashed lines describe the fluctuations for intermediate 
energies. The inset panel zooms  in on the region of minimal fluctuations at $x\simeq1$. 
}
\end{figure}

The statistical measure of fluctuations is  the standard deviation of shower particle number,
 $\sigma_N$. The corresponding values of  $\sigma_N(x,A,E)/N(x)$ obtained  from the
shower simulated dataset (Section III) are presented in Fig.~\ref{sigma} (symbols).
The inset panel shows the region of minimal fluctuations in detail.
The lines in Fig.~\ref{sigma} correspond to the parametrizations   
\begin{equation}
\frac{\sigma_N}{N}\simeq
\begin{cases}
a_1-a_2\ln{x}, &\text{if $x\leqq1$;}\\
a_1+a_3(\ln^{\eta}x)/x, &\text{if $x\geqq1$,}
\end{cases}
\end{equation}
where 
\[
\begin{split}
a_1& =0.165A^{-0.32}E^{-0.13}\;,\\
a_2& =0.68A^{-0.185}E^{-0.009}\;,\\
a_3&=3.77A^{-0.386}E^{-0.035}\;,\\
\eta&=2.67A^{-0.080}E^{-0.027}
\end{split} 
\]
at $\chi^2/470\simeq1.7$.

\section{Summary}
The standard inverse problem of cosmic ray physics in PeV-EeV energy region is the identification of
a primary nucleus (or elemental composition) and the estimation of its energy (or energy spectrum) by the detected
shower response at the observation level in the frames of a given interaction model.
The efficiencies of primary particle and primary energy estimators depend on both the accuracy (Section III) and
universality (Section V) of shower longitudinal profile description. 

Historically, the conventional shower longitudinal profiles were proposed in 1960 (Greisen function) \cite{Greisen},
1977 (Gaisser-Hillas function) \cite{GH} and 2001 (Gaussian-In-Age approach) \cite{HiRes1}. 
The efficiencies and accuracies of listed profile parametrizations are compared in Refs. \cite{Matthews,Song,Auger15} in detail. 

The last Gaussian-In-Age approach \cite{Matthews} reduced the number of parameters to 3, and 
decreased the intercorrelations between parameters of profile function in return for the narrow range of applicability in the vicinity of
 shower maximum: $0.75 \lesssim s \lesssim1.25$ \cite{Song}, where $s=3/(1+2/x)$ is the shower age parameters.\\
 
 The alternative shower longitudinal profile description (expressions 2-4), as opposed to the parametrizations 
 \cite{Matthews,GH,Greisen} , represents the first complete
 formula for shower profile, $N(T,A,E)$, depending on the
 atmosphere depth ($T$), primary nucleus ($A$) and primary energy $E$. 
 Expressions (2-4) provide the accuracies of about $2-3\%$ for the region of $0<T\le 1450$ g/cm$^2$, $A\le56$, $1 \text{ PeV}\le E\le10\text{ EeV}$.
The results are obtained in the frames of {\footnotesize{SIBYLL}} \cite{SIBYLL} interaction model (Section III).

The position of the shower maximum, $T_{max}(\boldsymbol{\varepsilon})$  from expression (4b) and profile shape function,
$\delta(x,\boldsymbol{\varepsilon})$ from expression (4a),  depend only on the primary nucleon energy $\boldsymbol{\varepsilon}=E/A$, 
which is in agreement with the prediction of superposition model \cite{Sommers}.

The amplitude of the profile  $N_{max}(E,\boldsymbol{\varepsilon})$ from expression (4c) depends on both the primary energy ($E$) 
and nucleon energy ($\boldsymbol{\varepsilon}$).

The intercorrelations between the 
$N_{max}(E,\boldsymbol{\varepsilon})$,  $T_{max}(\boldsymbol{\varepsilon})$ and $\delta(x,\boldsymbol{\varepsilon})$
shower profile parameters are negligible. 

The profile shape function, $\delta(x,\boldsymbol{\varepsilon}$), from (3) has the simple logarithmic
representation (5) which provides an analytic solution for the corresponding inverse profile function, which can
be used in the Constant-Intensity-Cut method \cite{GH}.

The fluctuations of particle shower longitudinal profile, $\sigma_N/N$, from parametrization (10) depend on
the energy ($E$) and mass number ($A$) of the primary nuclei.  
\section{Acknowledgments}
I wish to thank James Matthews for useful correspondence.

\end{document}